\documentclass{IEEEtran}
\usepackage{amsmath,amsfonts}
\usepackage{algorithmic}
\usepackage{algorithm}
\usepackage{array}
\usepackage[caption=false]{subfig}
\usepackage[justification=centering]{caption}
\usepackage{textcomp}
\usepackage{stfloats}
\usepackage{url}
\usepackage{verbatim}
\usepackage{graphicx}
\usepackage{xcolor}
\usepackage{cite}
\usepackage{tcolorbox}

\begin{document}

\title{Semantic and Goal-oriented Wireless \\Network Coverage: The Area of Effectiveness}

\author{Mattia Merluzzi,~\IEEEmembership{Member,~IEEE,} Giuseppe Di Poce, and Paolo Di Lorenzo,~\IEEEmembership{Senior Member,~IEEE}
        % <-this % stops a space
\thanks{M. Merluzzi and G. Di Poce are with Univ. Grenoble Alpes, CEA-Leti, France (e-mail: \{mattia.merluzzi, giuseppe.dipoce\}@cea.fr). \\
Paolo Di Lorenzo is with CNIT and Sapienza University of Rome, Italy
(e-mail: \{paolo.dilorenzo\}@uniroma1.it). }% <-this % stops a space
\thanks{This work was supported by the SNS JU project 6G-GOALS under the EU’s Horizon program Grant Agreement No 101139232, and the ANR under the France 2030 program, grant "NF-NAI: ANR-22-PEFT-0003", and by the European Union under the Italian National Recovery and Resilience Plan (NRRP) of NextGenerationEU, partnership on “Telecommunications of the Future” (PE00000001 - program “RESTART”).}\vspace{-.2cm}}



\maketitle

\begin{abstract}
Assessing wireless coverage is a fundamental task for public network operators and private deployments, whose goal is to guarantee quality of service across the network while minimizing material waste and energy consumption. These maps are usually built through ray tracing techniques and/or channel measurements that can be consequently translated into network Key Performance Indicators (KPIs), such as capacity or throughput. However, next generation networks (e.g., 6G) typically involve \textit{beyond communication resources}, towards services that require data transmission, but also processing (local and remote) to perform complex decision making in real time, with the best balance between performance, energy consumption, material waste, and privacy. In this paper, we introduce the novel concept of \textit{areas of effectiveness}, which goes beyond the legacy notion of coverage, towards one that takes into account capability of the network of offering edge Artificial Intelligence (AI)-related computation. We will show that radio coverage is a poor indicator of real system performance, depending on the application and the computing capabilities of network and devices. This opens new challenges in network planning, but also resource orchestration during operation to achieve the specific goal of communication.
\end{abstract}


\vspace{-.2cm}
\section{Introduction}
A typical and important task for a wireless network operator or an industrial player planning a private network is to estimate, given a deployment of wireless Access Points (APs), the coverage of such network within a given area. This can be based on the received signal strength (RSS) from (a subset of) deployed access points around the area \cite{Sarkar23}, or other metrics usually related to wireless performance, such as the capacity or throughput \cite{Jaziri2025}. These maps are useful to understand and assess whether there are coverage holes in the network, to be able to take countermeasures such as network densification or antenna improvements among others. Also, it is usually not feasible to measure the needed metrics (e.g., RSS) over the entire space of a network. Thus, one can resort on estimations and predictions based on sparse measurements \cite{Sarkar23,Almasan23,Sardellitti2024,Zaboub23}. The quality of such estimation depends on the location of available measurements, coupled with the topology of the environment and, obviously, the methodology used for estimation. For instance, a simple interpolation may fall short in the presence of blockages (e.g., a building) that abruptly decrease the received power. In this case, graph-based approaches can easily outperform classical interpolation methods \cite{Sardellitti2024, Almasan23}. 

To build a coverage map, we need the definition of a proper quality metric, which for wireless communications typically includes throughput, delay, resource block utilization, etc.
%\textbf{\textit{All these metrics pertain to wireless communication}}\\
However, next generation wireless networks not only promise great improvements in terms of communication performance, but also a tight integration of sensing, computing, and learning capabilities \textit{within the network}. This vision paves the way for sophisticated Artificial Intelligence (AI)-driven processing of the vast streams of data continuously collected from sensors, users, and machines, which will be essential for monitoring complex environments, making real-time decisions, and optimizing critical operations, including production processes and safety management \cite{Letaief2022}. To realize this vision, communication networks must evolve to accommodate new constraints and performance metrics associated with in-network AI-driven tasks. At the core of this evolution is the concept of in-network AI, which underpins the paradigm of edge intelligence (or \textit{edge AI}) that prioritizes localized data processing, reducing reliance on distant central clouds. This comes with several benefits including energy consumption, privacy, and backbone network load reduction, but also a few challenges concerning the management and orchestration of heterogeneous (e.g., communication, computation, and learning) network resources. Also, one fundamental objective is to keep such resources as limited as possible to avoid excessive resource usage/waste, while not sacrificing performance and/or values below acceptable levels. 

In this direction, the emerging paradigm of semantic and goal-oriented communications \cite{6GGOALS2023} will help designing procedures and algorithms to make intelligent agents at the edge cooperate efficiently and effectively, by only exchanging the information that is relevant for a task or meaningful for understanding. Crucially, new Key Performance Indicators (KPIs) that go beyond pure communication metrics appear, towards \textit{goal}- and \textit{semantic}-oriented metrics such as the effectiveness or the meaning of exchanged messages. In this context, the \textit{concept of coverage map must be rethought} to take into account the effectiveness (or, meaning conveying) that is guaranteed by a network deployment involving communication, but also computing resources/nodes embarked with more or less powerful AI models. This is fundamental when radio and computing resources co-exist and co-habit the same location, thanks to, e.g., edge/fog computing. Indeed, received signal strength and capacity metrics are insufficient to describe network coverage, plan new deployments, and orchestrate network resources. As a simple example, a user requesting an edge AI service from a computing node co-located with a wireless AP that provides excellent radio coverage may still experience poor quality from the edge inference service (or, more generally, from computing-intensive services), and vice versa. This depends on the network topology, the deployment of compute resources at wireless access points, and the AI models hosted on these computing nodes. Additionally, it is influenced by the contingent state of the network, including factors such as load and energy availability. 
%For example, a user in close proximity to a wireless access point typically experiences a high-capacity link. However, if the AP is equipped with a simple model capable of performing only low-accuracy tasks relative to the user’s requirements, or if the computing node is operating in energy-saving mode, the overall performance may be significantly degraded compared to expectations.
For example, a user near a wireless access point usually gets a high-capacity link. However, if the AP is equipped with a simple model capable of performing only low-accuracy tasks relative to the user’s requirements, or the computing node is in energy-saving mode, performance may fall short of expectations.\\
\noindent \textbf{Contributions.} The main objective of this paper is to redefine wireless network coverage, moving beyond its traditional definition toward a goal-oriented, semantic-aware design. We start reviewing the fundamentals of radio coverage maps and their construction, then highlight their shortcomings in the case of an edge classification service, where users require not just wireless connectivity but also computing and AI resources. For the first time in the literature, we introduce the novel concept of \textit{Area of Effectiveness (AoE),} incorporating the associated costs into a holistic framework that jointly considers communication, computing, and learning. We demonstrate the significant advantages of the proposed joint design for wireless network coverage over traditional approaches that separate communication and AI-driven computing. Additionally, leveraging these insights, we propose a set of network optimization strategies influenced by this redefined notion of coverage. Finally, we provide key recommendations that pave the way for future research challenges, driving network design and optimization toward semantic and goal-oriented communication tailored to evolving user needs.\\
\textbf{Outline.} The remainder of the paper is organized as follows. Sec. \ref{sec:sys_model} presents the system model and the relevant KPIs; Sec. \ref{sec:radio_coverage} recalls the basic concepts of pure radio coverage; Sec. \ref{sec:goeff_coverage} extends to semantic and goal-oriented coverage, introducing the proposed concept of area of effectiveness; finally, Sec. \ref{sec:recommendations} sets the recommendations and draw the conclusions.

\begin{figure*}
    \centering
    \includegraphics[width=0.95\textwidth]{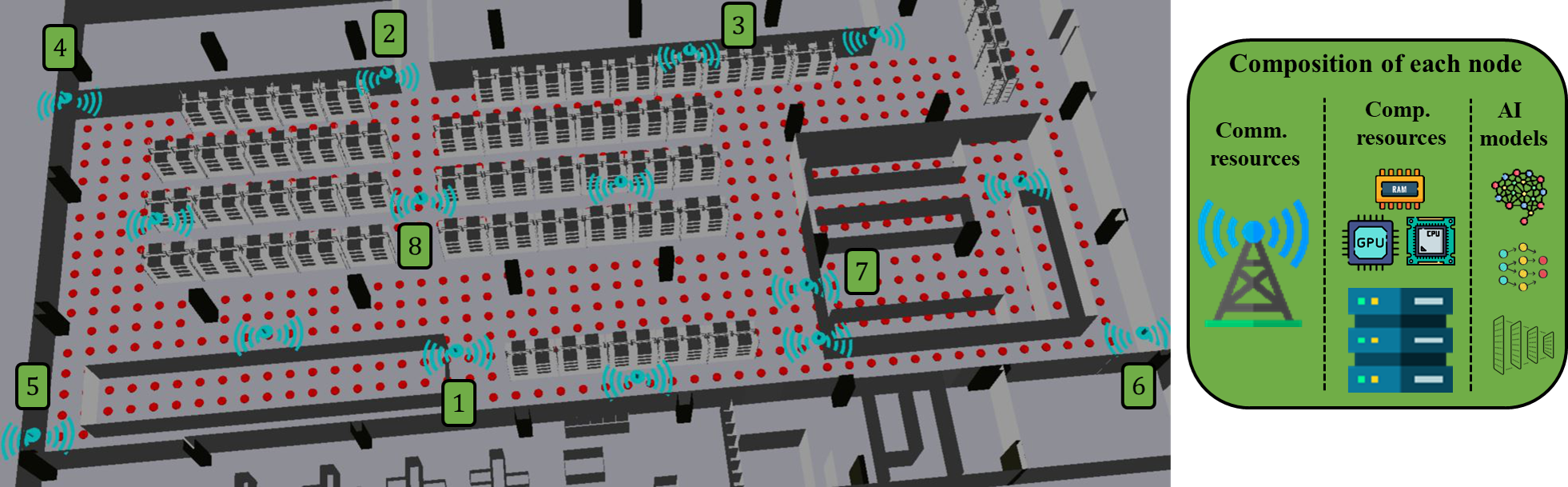}
    \caption{Indoor factory ray tracing scenario \cite{Jaziri2025} and AP node composition with radio computing and AI resources}
    \label{fig:scenario}
\end{figure*}

%\noindent \textbf{Contribution.} Within the described context involving such heterogeneous resources and services, this paper aims to promote a new way of defining the concept of coverage, to go beyond the legacy definition and overcome the limitations of the latter in next generation, semantic-enabled networks. 
%First, we recall the basic concepts of radio coverage maps and the necessary steps to build them. Then, we show the shortcomings of the latter for the simple case of an edge classification service, in which users not only need wireless, but also computing and AI resources. 

%We introduce the novel concept of \textit{\textbf{areas of effectiveness (AoE)}}, along with the associated costs in a holistic view that considers communication and computing. 
%We show how a joint approach is extremely beneficial, against approaches that treat communication and AI-related computing in a disjoint fashion. 

%Finally, based on the gained understanding, we propose a set of network optimization steps that will be impacted by this new definition of coverage along with some recommendations that will open new research challenges in tailoring network to semantic and goal-oriented communication.

\section{System model and KPIs}\label{sec:sys_model}
In this paper, we exploit the ray tracing data\footnote{The in-factory ray-tracing channel dataset was produced by Siradel using the Volcano software within frame of the POSEIDON (ANR-22-CE25-0015) and 5GSmartFact (MSCA No. 956670) projects.} from the perspective of radio coverage, complemented with specific models for computing performance (delay and inference performance). However, the concepts and methods easily apply to any network topology and deployment. In Fig. \ref{fig:scenario}, we illustrate the indoor deployment scenario presented in \cite{Jaziri2025}, comprising 15 APs placed at the indicated locations. Among all APs, for this study we consider the subset of $8$ APs indicated in the figure by the numbered green boxes. In this scenario, ray tracing has been performed for the locations indicated by the red dots, with squared pixels of side $2$ m. All APs operate at $f_c=3.7$ GHz, on a $360$ kHz bandwidth. As also illustrated in Fig. \ref{fig:scenario} (right hand side), we consider joint deployment of radio and computing resources. Namely, each AP is equipped with a computing unit, embarked with a set of pre-trained AI models ready to output inference results. The set of pre-trained models is possibly different across APs, with different performance and computational cost. A detailed set of parameters will be provided in Sec. \ref{sec:goeff_coverage}, along with performance evaluation. 
\subsection{Beyond communication KPIs}
In this work, we investigate edge inference as representative use case for goal-oriented semantic communication. For the latter, in the most simple case, an end device collects data from the environment (e.g., surveillance camera collecting video or any sensing device collecting signals), possibly pre-processes them to extract relevant features (i.e., semantics) and transmits these data (or, an intermediate representation) to an edge server that outputs a result with given effectiveness requirements (i.e., the goal). The latter could have different forms such as simply being a label, the detection of an object, or an action. Also, it can be based on data received by multiple sensors in a cooperative fashion. 
The objective of goal-oriented semantic communication is to \textit{tailor communication to the specific task or message interpretation capabilities}, in order to not waste communication resources. 

We explore different ways of building coverage maps, from the most classical ones including KPIs such as capacity, to new class that is defined on more elaborated KPIs. As proposed in \cite{Dilo23}, a goal is the fulfillment of a set of tasks (e.g., learning, control, actuation, etc.) characterized by a number of requirements that, if attained, determine its accomplishment. In the case of goal-oriented communication, a typical general KPI is the \textit{goal-effectiveness}, which can be defined as the probability for a system of achieving a desired goal, with the latter strongly depending on the allocated system resources in terms of communication and computation, as well as other exogenous environment states \cite{merluzzi20236g}. For instance, for the edge inference use case, the goal can be defined as \textit{receiving correct results within a deadline}, with the delay accounting for all phases, from sensing, to transmission, computing and actuation. Consequently, the goal-effectiveness is the probability that this event occurs under the time-varying environment states and system optimization. 
\begin{tcolorbox}
Differently from the capacity or data rate, this definition requires an \textbf{end-to-end perspective}, in which communication,  processing, and learning play a role. 
\end{tcolorbox}
In particular, not only the end-to-end delay is the accumulation of transmission and processing delays, but this also applies to energy consumption. Denoting by $n_b^i$ the application input bits (e.g., size of images from a camera), $R_l^u$ the uplink data rate, $w_p$ the number of FLOPS needed to process data with an AI model $m$, and $F$ the computing power (in FLOPS/s) allocated by the edge server, the end-to-end delay \cite{deSant21} is
\begin{equation}\label{loop_delay}
    D_{\text{loop}}=\frac{n_b^i}{R_l^u} + \frac{\omega_m}{F},
\end{equation}
where we neglected the delay for result transmission, which can be also integrated. The goal-effectiveness can be defined as in \cite{merluzzi20236g}, i.e., the probability of obtaining the correct result, with $D_{\text{loop}}\leq D_{\max}$, with $D_{\max}$ being a threshold required by the application. Note that radio ($R_l^u)$ and computing ($\omega_m$ and $F$) parameters usually evolve over time, making the loop delay a random variable.
Several other new KPIs have been defined in the context of semantic and goal-oriented communications. A subset of those is presented in \cite{Zhou24}, with the relation to the data type and a set of tasks. The objective of this paper is to show that the concept of coverage must evolve to take into account these KPIs in future network deployment and operations. To do so, we first briefly present a simple example of radio coverage, highlighting its shortcomings in the sense of semantic and goal-oriented communication.
\section{Radio coverage maps: a legacy concept}\label{sec:radio_coverage}
Typically, given a network deployment and a propagation environment (e.g., indoor or outdoor), a radio coverage map is built based on the capability of the deployed wireless APs to serve a set of locations in space. In the most simplistic case, each location in space is characterized by the received power (or, Received Signal Strength - RSS) by each AP. By selecting the strongest RSS among all APs (or, a set of them in cell-free deployments \cite{Jaziri2025}, a map of the RSS or the resulting capacity (easily computable via the Shannon formula) can be constructed, to identify possible coverage holes. One way to build such maps is to resort on ray tracing techniques, which simulate reflections and diffractions of electromagnetic waves in a given environment.
\begin{figure}
    \centering
    \includegraphics[width=\columnwidth]{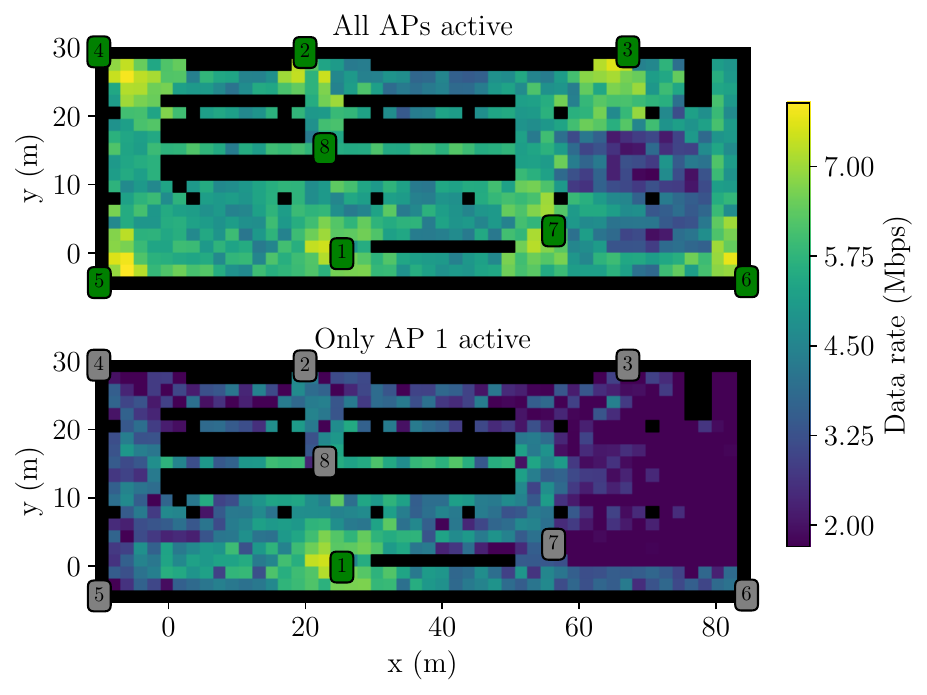}
    \caption{Typical radio-based coverage maps}
    \label{fig:capacity_cov}
\end{figure}
Then, the considered area is divided in equally sized pixels, and a quality metric (e.g., the capacity or spectral efficiency) is associated to each pixel. To correctly assign a quality metric in the presence of multiple APs, an association policy must be defined, with the latter being a rule to associate a given location (pixel) to an AP. As a simple example, a capacity-based radio coverage map is constructed by assigning each location to the AP that guarantees the strongest RSS. In this way, each location is guaranteed to experience the best radio signal quality. Obviously, such a procedure does not take into account possible contextual information typical of network operations, such as an anomalous density of users in a particular area, which could change the actual performance due to spectrum resource sharing. In this paper, we only consider per-location coverage as a deployment problem, without considering contingent network occupancy in case of multiple users during operation. We leave this aspect to future investigations in which coverage maps might evolve over time, according to user density, energy availability, etc. 

In this section, we show a first radio coverage map, with the most classical association policy, which assigns a given location to the AP providing the highest RSS. Considering the scenario illustrated in Section \ref{sec:sys_model} (cf. Fig. \ref{fig:scenario}), we show in Fig. \ref{fig:capacity_cov} the coverage in terms of achievable data rate (in Mbits/s - using the Shannon formula for the capacity) in two different conditions: \textit{i)} when assuming all APs are available (first map), and \textit{ii)} when only AP $1$ is available (second map). Available APs are in green, while unavailable ones in gray color.
Obviously, the best coverage is obtained in the scenario with all APs available, as for each location the best channel can be opportunistically selected via the RSS-based association policy. However, coverage holes are not entirely avoidable, e.g., the blue area on the right hand side of the figure (corresponding to the room within walls). The black areas denote the racks and walls visible in Fig. \ref{fig:scenario}. Looking at these maps, with no surprises, it is clear that the RSS-based policy is the best possible solution to guarantee radio coverage across space. In other words, \textbf{the association policy and the quality metric used to build the map match perfectly}, which makes the current methods for deployment fully adapted to legacy communication systems. 
\begin{tcolorbox}
\textbf{\textit{What if this last condition falls short, i.e., the (goal-oriented) KPIs mismatch the association policy?}}
\end{tcolorbox}
In Section \ref{sec:goeff_coverage}, we give an answer to this question for the case of edge inference. We show that the maps illustrated in Fig. \ref{fig:capacity_cov} bring very little information about the requirements related to semantic and goal-oriented communication. We argue and motivate that the concept of coverage itself, as well as the policies to assign locations to serving APs, need a redefinition that takes into account effectiveness requirements.

\section{Goal and semantic-aware coverage maps}\label{sec:goeff_coverage}
We now demonstrate how \textit{communication and computing matter} when assessing network coverage.  First, we highlight the limitations of pure radio coverage maps by introducing AI-based in-network computing. To illustrate this, we analyze the same scenario while incorporating all the resources depicted in Fig. \ref{fig:scenario} (right-hand side).

We focus on edge classification as a service. Among the various possible KPIs, we focus on goal-effectiveness, which is defined as the percentage of samples correctly classified within a given deadline (cf. \eqref{loop_delay}). We assume that each AP is equipped with a computing unit and single classification model, pre-trained on the \textit{imagenette} dataset \cite{Howard_Imagenette_2019}. Each model $m$ comes with performance in terms of accuracy, and cost in terms of computing load $\omega_m$ (cf. \eqref{loop_delay})\footnote{\url{https://pytorch.org/vision/0.20/models.html}}:
\begin{itemize}
    \item \textit{MobileNetv3}: $\omega_m=0.11$ GFLOPS, $\alpha_m=0.957$;
    \item \textit{Resnet-50}: $\omega_m=8.17$ GFLOPS, $\alpha_m=0.9858$;
    \item \textit{Resnet-101}: $\omega_m=15.5$ GFLOPS, $\alpha_m=0.989$;
    \item \textit{vit\_b\_16}: $\omega_m=33.6$ GFLOPS, $\alpha_m=0.996$.
\end{itemize}

We use the concept of loop delay (cf.\eqref{loop_delay}) and set a deadline $D_{\max}=0.5$ s. Given an association policy, each location experiences a transmission delay (first term in \eqref{loop_delay}). Then, depending on the available model at the selected AP, computing resources ($F$ in FLOPS/s) are allocated to meet the deadline. If the latter exceed $F_{\max}=1$ TFLOPS/s, a failure occurs. Obviously, a more powerful model helps in terms of accuracy, but requires more computing resources, thus could fail due to the loop delay. To show the drawbacks of a disjoint system optimization and the need for a holistic view of resources, we consider the three following association policies:
\begin{itemize}
    \item \textbf{\textit{RSS-based association:}} Each location is assigned to the AP that provides the highest RSS. Since all APs offer the same bandwidth, this approach also serves as a capacity-based association.
    \item \textbf{\textit{Model-based association:}} Each location is always assigned to the AP embedded with the best model in terms of accuracy (in this case, \textit{vit\_b\_16}). 
    \item \textbf{\textit{Genie-aided association:}} In a dynamic manner, each location is assigned to the access point (AP) that delivers the correct result within the deadline for each data sample, while minimizing the computing load (measured in GFLOPS/s). If no model produces a correct result, a goal failure occurs. This represents an upper-bound on performance, assuming a priori knowledge of model outputs for each data sample.
\end{itemize}

\begin{figure*}[t]
    \centering
    \subfloat[Goal-aware coverage maps]{
        \includegraphics[width=\columnwidth]{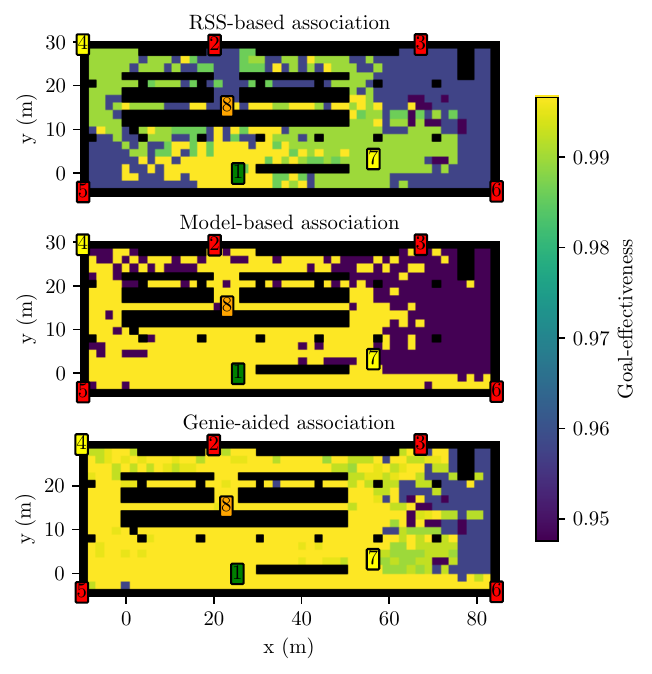}
        \label{fig:goeff_map}
    }
    \subfloat[Area of Effectiveness with threshold $q_{\text{th}}=0.99$]{
        \includegraphics[width=0.8\columnwidth]{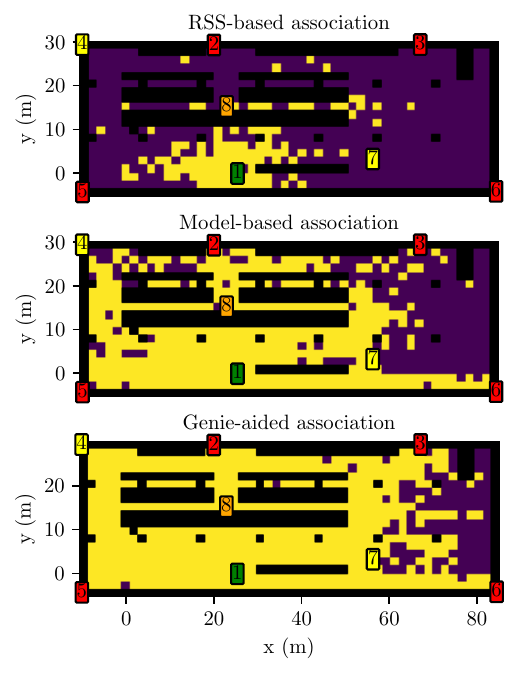}
        \label{fig:aoe}
    }
    \caption{Goal-aware coverage maps (a) and area of effectiveness (b)}
    \label{fig:goeff_coverage}
\end{figure*}
%\begin{figure}[t]
 %   \centering
 %   \includegraphics[width=\columnwidth]{goeff_coverage.pdf}
 %   \caption{Goal-aware coverage maps and area of effectiveness with threshold $q_{\text{th}}=0.99$}
  %  \label{fig:goeff_coverage}
%\end{figure}
\subsection{Areas of effectiveness}
In this section, we introduce a new concept, which we refer to as \textit{Area of Effectiveness (AoE)}, and is defined as follows. \smallskip 
\begin{tcolorbox}
\textbf{Area of Effectiveness:} It is a portion of physical space $\mathcal{S}$ where the goal-effectiveness $\mathcal{E}_{go}$ is greater than a threshold $q_{th}$, given a target application and a deployment of connect-compute network resources. Formally, the AoE can be defined as follows:
\begin{equation}
    {\rm AoE} = \{ \mathcal{S}\,|\, \mathcal{E}_{go}(\mathcal{S}) \geq q_{th}\}. \nonumber
\end{equation}
Given an area $\mathcal{A}$ of investigation, we have ${\rm AoE}\subseteq \mathcal{A}$.
\end{tcolorbox}
Given a deployment scenario, as well as time-varying parameters (such as energy availability, user density, etc.), different AoEs may emerge within the same environment. These areas strongly depend on the association policies adopted by the system during operation. As an example, in Fig. \ref{fig:goeff_coverage}, we illustrate the goal-aware coverage maps (Fig. \ref{fig:goeff_map}) and an example of AoE (Fig. \ref{fig:aoe}), for the described scenario, considering different association policies. In particular, while Fig. \ref{fig:goeff_map} shows the actual effectiveness around the scenario, Fig. \ref{fig:aoe} is obtained by setting a threshold $q_{\text{th}}=0.99$ for the effectiveness. Then, by construction, the latter is a binary map in which the effectiveness constraint is either met (yellow area, i.e., the AoE) or not (blue area). APs are indicated in different colors depending on the deployed inference model: red (\textit{MobileNetv3}), orange (\textit{Resnet-50}), yellow (\textit{Resnet-101}) and green (\textit{vit\_b\_16}). The first map is the goal-effectiveness map obtained with the \textit{RSS-based association}. 
%Obviously, the best effectiveness area is the one around AP $1$, i.e., the one embedded with the vision transformer. 
%In a general way, the goal-effectiveness in the considered scenario directly comes from the model that present at the AP that is selected only according to radio parameters. 
Naturally, looking at Fig.\ref{fig:goeff_map}, the highest effectiveness is observed in the area surrounding AP $1$, which is equipped with the vision transformer. More generally, in the given scenario, goal-effectiveness is determined by the model available at the AP selected solely based on radio parameters. We can already argue that Fig. \ref{fig:goeff_coverage} is significantly more informative than Fig. \ref{fig:capacity_cov} from a goal-oriented perspective. Furthermore, a user located far from AP $1$ \textbf{would not achieve the highest accuracy under a purely radio-based association.} In general, a user placed close to an AP with a poorer inference model in terms of accuracy, will always get the performance of the latter. Moving the attention to the \textit{model-based association} policy (center maps in Fig. \ref{fig:goeff_coverage}), we can notice that the latter comes with a binary behavior in terms of coverage, i.e., either a location is served with the best goal-effectiveness (thanks to \textit{vit\_b\_16} deployed in AP $1$) or it is not served at all (i.e., goal-effectiveness $0$) due to its \textbf{excessive distance (in the radio propagation sense) from the best model}. In other words, the transmission delay is too severe to reach the best model within the deadline, and with a constrained computational power $F_{\max}$. From Fig. \ref{fig:goeff_coverage}, we can notice that the locations on the right hand side of the scenario cannot be served at all. However, all other locations experience the highest goal-effectiveness. These first two associations are clearly sub-optimal, as they treat communication and computation aspects \textit{disjointly}.
As a third association policy, let us now focus on the best performance that a location could get with the \textit{genie-aided association}. In this case, we can notice that all areas in which the model-based association fails can be actually covered, with the exception of a few examples. Obviously the obtained goal-effectiveness is not the highest in all locations, as the \textit{genie takes into account accuracy and loop delay when selecting the best AP}. In this way, the non-covered areas by the \textit{model-based association} are covered with a goal-effectiveness that depends on which model is capable of offering the service under the delay constraints, and with the lowest computing resources. Finally, Fig. \ref{fig:aoe} shows how the genie-aided joint strategy helps extending the AoE with respect to disjoint methods, for the selected threshold $q_{\text{th}}=0.99$.

% From the understanding we gain from Fig. \ref{fig:goeff_coverage}, we can define a new concept, which we name as \textbf{\textit{area of effectiveness (AoE)}}. An AoE is a portion of physical space in which the goal-effectiveness is positive. Given a deployment scenarios, but also time-varying parameters (including energy availability, user density, etc.) different AoEs can result in a given environment. The latter strongly depend on the association policies adopted by the system during operation.

%\textbf{\textit{Remark.}} \textit{The areas of effectiveness as presented in this section do not bring any information about the cost incurred by the system, e.g., in terms of communication and computation.}

\subsection{The costs of effectiveness}
\begin{figure*}[t]
    \centering
    \subfloat[Computational load map]{
        \includegraphics[width=0.98\columnwidth]{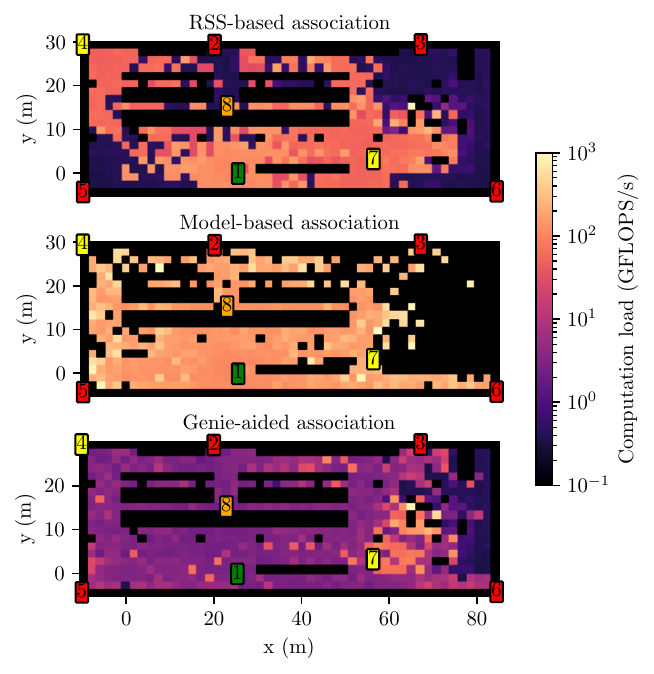}
        \label{fig:comp_load_map}
    }
    \subfloat[AP activity time map]{
        \includegraphics[width=0.98\columnwidth]{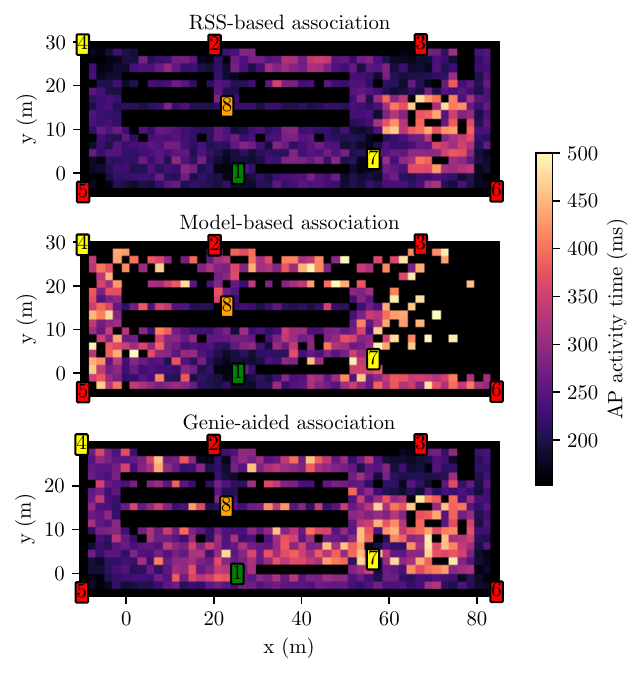}
        \label{fig:delay_map}
    }
    \caption{Computational load and AP activity maps as the costs of effectiveness}
    \label{fig:cost_maps}
\end{figure*}
The areas of effectiveness as presented in the previous section do not bring any information about the cost incurred by the system, e.g., in terms of communication and computation. Thus, it becomes fundamental to assess what is the cost, under a given association policy, to achieve the goal-effectiveness illustrated in Fig. \ref{fig:goeff_coverage}. As such, we build two other maps for communication and computation costs:
\begin{itemize}
    \item \textbf{\textit{Computational load map:}} For a given location in space, this represents a cost map in terms of computing power (measured in GFLOPS/s) incurred by the network under a specific association policy. Other metrics, such as memory usage, could also be taken into account.
    \item \textbf{\textit{AP activity map:}} It represents the activity time map (in milliseconds) required by the AP to serve a location under a specific association policy. This metric is closely linked to energy consumption, particularly when APs can enter low-power sleep modes  during periods of inactivity \cite{Salem19}.
\end{itemize}
The computational load map for all association policies is presented in Fig. \ref{fig:comp_load_map}, following the same order as in Fig. \ref{fig:goeff_coverage} for \textit{RSS-based}, \textit{model-based}, and \textit{genie-aided} association. As we can notice from Fig. \ref{fig:comp_load_map}, the \textit{model-based association} (center map) comes with the highest possible cost in terms of computing across the considered scenario. This is because each location is assigned to AP $1$, which not only classifies with the most resource consuming model, but is also far from several locations. Indeed, the computation cost increases with distance due to the longer communication delay, which must be accounted for to meet the loop delay constraint (cf. \eqref{loop_delay}). Naturally, the computation cost is zero outside the areas of effectiveness. The \textit{RSS-based association }(top map) comes with a lower cost in locations that are closer to the APs, especially in the case of a simpler model (e.g., AP $5$ being embedded with \textit{MobileNetv3}). Obviously, the cost is the highest around AP $1$, as all locations around that area will exploit \textit{vit\_b\_16} for classification. Both disjoint strategies come again with clear drawbacks, which also include computational costs. Interestingly the \textit{genie-aided association} (bottom map), which jointly consider communication and computation aspects results in the lowest possible computational load overall across the scenario. The highest load is experienced by the areas within the room (cf. Fig. \ref{fig:scenario} and Fig. \ref{fig:capacity_cov}, right hand side), as they experience bad channel conditions with all APs. 

From a communication perspective, Fig. \ref{fig:delay_map} illustrates the communication cost (represented by the APs' activity time) incurred by the network to serve different locations under the three association policies. As expected, the \textit{RSS-based association} achieves the lowest cost, due to the proximity of APs in terms of radio coverage. The \textit{model-based association} represents the worst case on average, due to the furthest locations. Finally, the \textit{genie-aided association} comes with intermediate cost due to the way it is built, which inherently refuses bad channel conditions due to the delay constraint.

\begin{figure*}[t]
    \centering
    \subfloat[Computational load related metric]{
        \includegraphics[width=0.95\columnwidth]{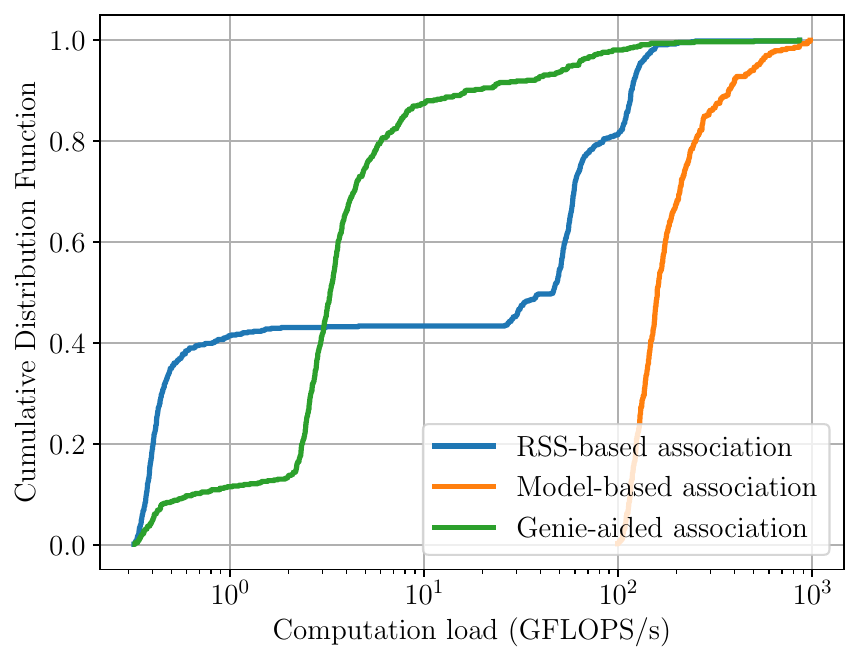}
        \label{fig:cdf_comp_load}
    }
    \subfloat[Communication load related metric]{
        \includegraphics[width=0.95\columnwidth]{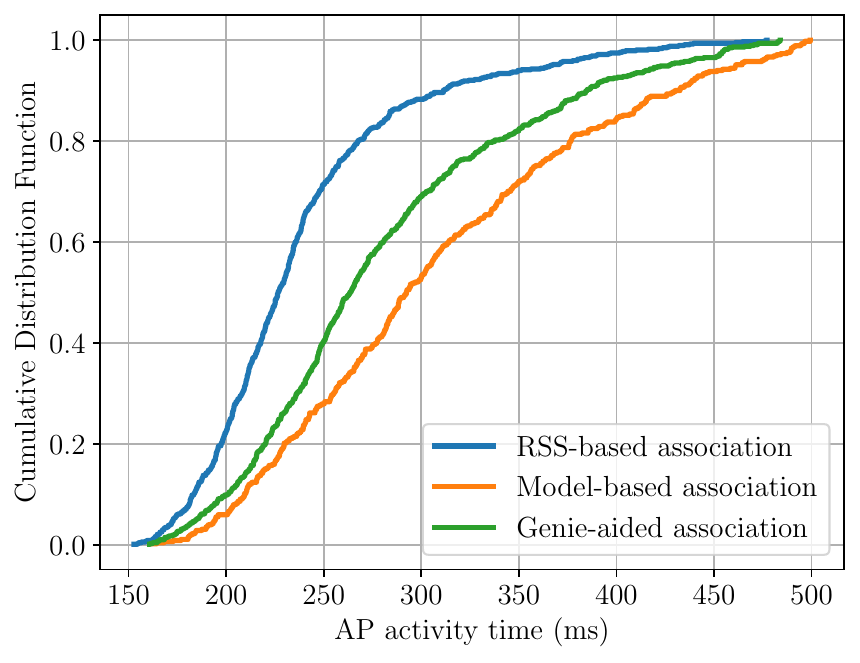}
        \label{fig:cdf_ap_activity}
    }
    \caption{CDF of the computational load and the AP activity time for the given scenario.}
    \label{fig:cdfs}
\end{figure*}

Another visualization of this result is presented in Fig. \ref{fig:cdfs}, where we plot the cumulative distribution function (CDF) of the computational load (Fig. \ref{fig:cdf_comp_load}) and the AP activity time (Fig. \ref{fig:cdf_ap_activity}), respectively, across the whole scenario. Looking at Fig. \ref{fig:cdf_comp_load}, the \textit{model-based association} experiences the worst performance, due to the systematic selection of AP $1$, which can be far (in the electromagnetic sense) from several locations, and also is embarked with the most computationally hungry inference model. The \textit{RSS-based association} experiences the highest percentage of computational load below 2 GFLOPS/s, benefiting from areas surrounding the various APs, which experience very low communication delay and, consequently, a more relaxed computing delay (cf. \eqref{loop_delay}). In contrast, the \textit{genie-aided association} policy achieves the lowest average cost and the highest percentage of workloads below 10 GFLOPS/s. This advantage stems from its ability to select the optimal AP that minimizes computation cost while meeting the performance goal. Finally, Fig. \ref{fig:cdf_ap_activity} presents similar results, with the exception that the \textit{RSS-based association} achieves the best performance in terms of communication, as it optimally selects APs from a radio perspective. However, this advantage comes at the cost of reduced effectiveness and lower computation savings. Despite the fact of being an upper bound, these results suggest that a \textbf{joint connect-compute approach not only helps extending the AoE, but also reducing the cost for the network in terms of computation}. 
\begin{tcolorbox}
Overall, network deployment and orchestration should jointly consider all aspects of wireless communication and computation, \textbf{taking into account both the AoE and the associated costs}. 
\end{tcolorbox}
Achieving an optimal balance between communication and computation depends on the hardware used for each function.

\section{Conclusions, new challenges and recommendations for network optimization}\label{sec:recommendations}
In this paper, we presented the new concept of area of effectiveness, i.e., an evolved way of defining wireless network coverage for semantic and goal-oriented communication. The results presented in the previous sections, despite being related to a use case and solely for the case of goal-oriented edge inference, open new challenges in network planning and dynamic orchestration. Our first claim is that \textit{\textbf{computing and AI-related aspects must be taken into account during network planning and coverage assessment}}. Obviously, the planning must also take into account the costs, which go beyond pure communication infrastructure costs, with new hardware dedicated to AI processing. Once the network is deployed, operations must be optimized by taking into account dynamic aspects related to communication and computing, with association policies that consider semantic and goal-oriented aspects.When the assignment is performed by the network in an agnostic manner, user mobility and path planning enable movement through areas of effectiveness tailored to the service. This has significant implications for various use cases, including autonomous vehicles and collaborative robots \cite{Kai24}.

Another important aspect is local semantic extraction, which, when computing resources are available, enables the transmission of only relevant information. This can enhance coverage in areas with low RSS by leveraging the benefits of compression and increased robustness.
%Other aspects pertain to local semantic extraction, in case of available computing resources, to transmit only the relevant information and possibly extend the coverage to areas with low RSS, thanks the gained compression and robustness. 
A simple way is to exploit deep neural network splitting, which we did not presented in this paper due to the lack of space. In this context, the issue of semantic model misalignment is crucial for evaluating coverage from the perspective of agents' mutual understanding \cite{Sana2023}. Finally, in this paper we considered only static maps, without taking into account network state variations such as user density and energy availability. Future investigations include the dynamic nature of networks, with coverage and areas of effectiveness that evolve over time, calling for continuous adaptation of user association, path planning and resource allocation, based on an augmented information that entails communication and computing aspects.
\section*{Acknowledgments}
The authors would like to thank Dr. David Demmer and Dr. Jean-Baptiste Doré for their support in extracting the exploited ray-tracing data by \textit{Siradel} introduced in Section \ref{sec:sys_model}.
\bibliographystyle{IEEEtran}
\bibliography{Main}
%\newpage
\section{Biographies}
\vspace{-2.5 cm}
\begin{IEEEbiography}[{\includegraphics[width=1in,height=1.25in,clip,keepaspectratio]{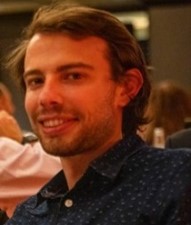}}]{Mattia Merluzzi} received the Ph.D. degree in Information and Communication Technologies from Sapienza University of Rome, Italy, in 2021. He is currently a research engineer at CEA-Leti, France, where he is involved in the research activities on 6G. He is actively participating in the SNS 6G-GOALS project, and he is the principal investigator for CEA in SUSTAIN-6G, the European lighthouse project on 6G and sustainability. He has participated in the H2020 European 6G flagship project Hexa-X, the H2020 EU/Japan project 5G-Miedge, the H2020 EU/Taiwan project 5G CONNI, and different national research projects in Italy and France. His primary research interests are in beyond 5G systems, stochastic network optimization, and edge machine learning/artificial intelligence. He was recipient of the 2021 GTTI Best Ph.D. Thesis Award. 
\end{IEEEbiography}
\vspace{-2.5 cm}
\begin{IEEEbiography}[{\includegraphics[width=1in,height=1.25in,clip,keepaspectratio]{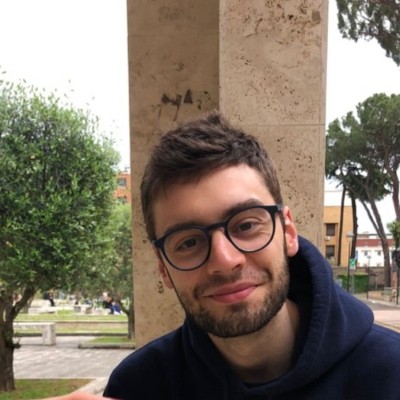}}]{Giuseppe Di Poce} is a Ph.D. student at CEA-Leti in Grenoble, France. He holds a Master of Science in Data Science from Sapienza Università di Roma, where he graduated with 110/110 cum laude with a thesis about Bio inspired Federated Learning and distributed non convex optimization.
Giuseppe has gained valuable experience as a Computer Science Teacher and as a Teaching Assistant in Probability Calculus and Statistics at Sapienza University. His expertise spans signal processing and machine learning, with hands-on experience from programs like the AWS Training Camp at Amazon Web Services.
Beyond academia, he is an active contributor to open-source projects, regularly sharing insights and research on GitHub. Passionate about data science, AI, Giuseppe continues to expand his knowledge and professional network in the tech community. 
\end{IEEEbiography}
\vspace{-2.5 cm}
\begin{IEEEbiography}[{\includegraphics[width=1in,height=1.25in,clip,keepaspectratio]{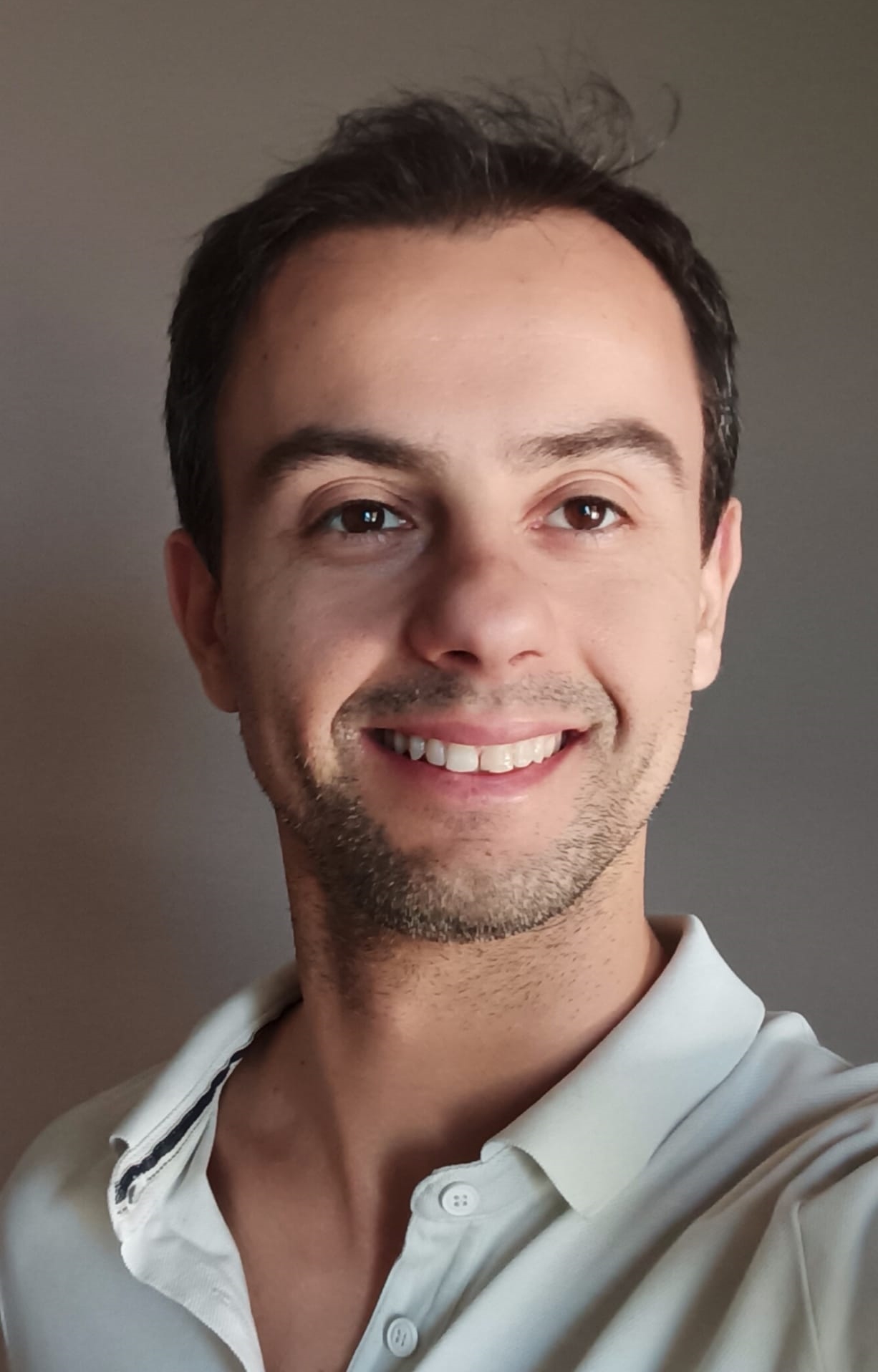}}]{Paolo Di Lorenzo} (Senior Member, IEEE) received the M.Sc. and Ph.D. degrees in telecommunication engineering from the Sapienza University of Rome, Italy, in 2008 and 2012, respectively. Currently, he is an Associate Professor with the Department of Information Engineering, Electronics, and Telecommunications at Sapienza University of Rome, Rome, Italy. He is the technical manager of the SNS-JU European Project 6G-GOALS, and was the Principal Investigator of CNIT-Sapienza Research Unit in the H2020 European Project RISE 6G. His research interests include topological signal processing, goal-oriented and semantic communications, distributed optimization, and federated learning. He held a visiting research appointment with the Department of Electrical Engineering, University of California at Los Angeles, Los Angeles, CA, USA. He is the recipient of the 2022 EURASIP Early Career Award, and of three best student paper awards at IEEE SPAWC10, EURASIP EUSIPCO11, and IEEE CAMSAP11, respectively. He is also the recipient of the 2012 GTTI (Italian National Group on Telecommunications and Information Theory) Award for the Best Ph.D. Thesis in communication engineering. He served as an Associate Editor for the IEEE Transactions on Signal Processing, for the IEEE Transactions on Signal and Information Processing over Networks, and for the EURASIP Journal on Advances in Signal Processing. He is currently a Senior Area Editor of the IEEE Transactions on Signal Processing.
\end{IEEEbiography}
\end{document}